\documentclass{article}
\usepackage[final]{neurips_2024}
\usepackage[utf8]{inputenc} 
\usepackage[T1]{fontenc}    
\usepackage{hyperref}       
\usepackage{url}            
\usepackage{booktabs}       
\usepackage{amsfonts}       
\usepackage{nicefrac}       
\usepackage{microtype}      
\usepackage{xcolor}         
\usepackage{hyperref}
\usepackage{graphicx}

\title{Powering LLM Regulation through Data: Bridging the Gap from Compute Thresholds to Customer Experiences}
\author{%
  Wesley Pasfield\\
  xD, U.S Census Bureau\\
  Washington D.C\\
  \texttt{william.w.pasfield@census.gov}
}
\date{August 2024}

\begin{document}

\maketitle

\begin{abstract}
    The rapid advancement of Large Language Models (LLMs) has created a critical gap in consumer protection due to the lack of standardized certification processes for LLM-powered Artificial Intelligence (AI) systems. This paper argues that current regulatory approaches, which focus on compute-level thresholds and generalized model evaluations, are insufficient to ensure the safety and effectiveness of specific LLM-based user experiences. We propose a shift towards a certification process centered on actual user-facing experiences and the curation of high-quality datasets for evaluation. This approach offers several benefits: it drives consumer confidence in AI system performance, enables businesses to demonstrate the credibility of their products, and allows regulators to focus on direct consumer protection. The paper outlines a potential certification workflow, emphasizing the importance of domain-specific datasets and expert evaluation. By repositioning data as the strategic center of regulatory efforts, this framework aims to address the challenges posed by the probabilistic nature of AI systems and the rapid pace of technological advancement. This shift in regulatory focus has the potential to foster innovation while ensuring responsible AI development, ultimately benefiting consumers, businesses, and government entities alike.
\end{abstract}

\section{Introduction}

This paper aims to motivate two key concepts. First, there is no standardized certification process to ensure large language model (LLM) powered  artificial intelligence (AI) systems are able to perform at an acceptable level compared to certified human experts at domain-specific tasks, leaving a critical gap in consumer protection. 

Second, while dynamic compute thresholds could serve as a means to trigger model registration and transparency requirements, they are not a sufficient means to reflect model risk \cite{hooker2024compute}. Domain-specific datasets should be the engine that powers LLM regulation for the purposes of consumer protection.   

LLM performance evaluation is unsettled, and led by industry and academia. Existing model benchmarks are helpful to distinguish the general efficacy of different models but are often irrelevant to specific use cases \cite{biyani2024domain}.

Research efforts \cite{raju2024judge, biyani2024domain} and guidance \cite{nistai} have emphasized the need for domain-specific testing, but proposed government regulation and intervention have been focused at the compute level, in part due to lack of access to customer use cases \cite{nicholas2024Policy}. Executive Order (EO) 14110 set a compute threshold as a means for regulation \cite{exec14110}, and the now-vetoed California State SB 1047 bill \cite{sb1047} also used the same threshold to define “covered” models subject to liability from catastrophic harms, including  from derivative models, leading to significant push-back from industry \cite{anthropic2024sb1047, meta2024response, openai2024response}. The EU AI Act \cite{euact} uses also uses a fixed threshold to identify general-purpose AI models that could cause systemic harm. Other regulatory suggestions \cite{kenny2023regulatable}, involve model explainability, which is certainly a promising research area \cite{templeton2024monosemanticity}, but is still evolving from a technological perspective for LLM-based techniques. 

Compute-level regulation is indirectly related to consumer experiences, as the underlying LLM is just a single component of a full LLM-based user experience, and compute and model performance are not perfectly correlated \cite{hooker2024compute}. The prompt, additional data and context provided by downstream developers, in addition to model fine-tuning and optimizations, have a significant impact in shaping the content that users receive.

Testing AI systems for functional sufficiency should be viewed as an essential complement to algorithmic assessments for fairness. A practicing mental health professional would face consequences if they did not maintain their required license according to the relevant state or federal law, and importantly would have a clear and universally understood path to obtain that license. If a company wanted to create an LLM-based mental health coach, there are some existing regulatory frameworks, like the FDA regulation of medical devices \cite{FDAAIRegFramework, FDAMobileMedicalApplications}, however the means of an LLM-based product evaluation to evaluate "clinical proficiency" at peer level with (human) licensed mental health professionals is not clearly established \cite{mesko2023healthcare, raza2024healthcare}.

Combining the two concepts of compute-level regulation and generalized model evaluation, consumers have little insight into the accuracy of domain-specific LLM-based experiences, and companies have no clear standardized approach to prove that their LLM-based experiences are safe and effective. The primary purpose of this paper is to highlight the regulatory, data and evaluation gaps that exist with LLM-based experiences, not to be overly prescriptive on the tactical solution, or the governing body responsible for the certification process.

\section{Background}
\subsection{Existing Proposed AI Regulation}

Recent  AI advancements driven by LLMs have led to more calls for government regulation to ensure the responsible development and delivery of the LLM-based experiences to consumers \cite{mesko2023healthcare, raza2024healthcare}. 

The probabilistic and open-ended nature of AI-based solutions makes them extremely difficult to regulate and control, putting regulators directly between two opposing motivations. Regulate too aggressively, and it risks stifling innovation of a potentially transformative technology. Regulate too conservatively, and it risks causing significant harm to consumers of the technology and erosion of consumer trust in LLM-based system output. 

Regulation to date has focused on the compute-level. Compute-level regulations have three clear limitations as the primary mechanism of consumer protection. First, fixed compute thresholds are brittle and quickly become outdated. Better data quality and improved model optimization and architectures can drive model performance increases, obfuscating the relationship between compute and performance \cite{hooker2024compute}. The increased proliferation of "small" language models, and adoption of techniques like distillation that allow for more efficient learning, further threatens the validity of the existing fixed thresholds proposed \cite{lambertregulation, hooker2024compute}.

Second, primarily focusing regulation and accountability mechanisms on foundational model developers rather than downstream systems stifles competition, innovation, and open source, as only the wealthiest developers will be able to take the associated financial risk \cite{lambertregulation, meta2024response}.  

Third, compute-level regulation is not directly related to the end-user experience. LLM-based experiences combine model outputs with other relevant contextual information to present information back to end consumers. A compute threshold serves as a proxy for a model's \textit{potential} capabilities. As a crude analogy, this amounts to the FDA only evaluating the independent ingredients that go into making a drug, rather than the drug itself. 

California Governor Gavin Newsom's SB 1047 veto message \cite{newsom2024veto} articulates the limitations of compute-level regulation as the primary means of consumer protection. "By focusing only on the most expensive and large-scale models, SB 1047 establishes a regulatory framework that could give the public a false sense of security about controlling this fast-moving technology. Smaller, specialized models may emerge as equally or even more dangerous than the models targeted by SB 1047."

\subsection{Existing LLM Evaluation Limitations}

The limitations associated with compute-level regulation and lack of specificity in evaluation of LLM-based applications is an especially acute issue in application areas that deal with sensitive or high-risk scenarios. This paper does not attempt to define high-risk\cite{euact, coloradoact}, just uses the term to acknowledge that these types of use cases exist.

Existing LLM evaluation gravitates towards generalized model performance. Examples such as Hugging Face open leaderboards \cite{huggingface} and Stanford's HELM \cite{Liang2022Helm} show an aggregation of performance across a variety of datasets. Increases in performance on generalized benchmarks are likely correlated with increases in performance on specific experiences, but the further away a use case is from the datasets used in calculating generalized benchmarks, the less certain that correlation becomes. Domain-specific evaluation datasets give a more direct view on experience efficacy. \cite{guo2024synthetic, raju2024judge, biyani2024domain}. 

Some models have successfully passed certification tests like the Uniform Bar Exam and the US Medical Licensing Exam \cite{achiam2023gpt4, singhal2023usmle}. To date, we have not observed any advocates for using the ability to pass these human-focused exams as a means for algorithmic certification.

It is common parlance to reference “vibes-based" evaluation as a means to evaluate LLM performance \cite{ruder2024Eval}. The Chatbot Arena is a great example of a quantified version of this type of evaluation, as it crowd-sources model performance based on human ranking of model performance \cite{Chatbotarena}. This quantitative, crowd-sourced approach is helpful for lower risk customer experiences, but higher risk scenarios require a more rigorous approach \cite{hung2023tightrope, mesko2023healthcare, raza2024healthcare, Bavaresco2024Judge}. 

\section{Proposed Certification Process}

\subsection{Certification Motivation}

The combination of curated and maintained datasets, as well as a certification process, promises significant benefits for consumers, businesses and certifying authorities.

For consumers the benefits are twofold. First, rather than relying on generalized model performance and compute thresholds as the sole representations of model efficacy, which both serve as a proxy for \textit{all} model capabilities, a certification process enabled by the curation of use-case specific datasets would give consumers confidence that their \textit{specific} LLM-based interactions are safe and effective. 

Second, it enables experiences that businesses may deem too risky on the surface to be solved through an LLM-based approach. In areas like healthcare, this could democratize care, particularly in hard-to-serve areas. Both of these benefits are directly referenced in EO 14110 Section 8 \cite{exec14110}.

For businesses, it offers a clear path to credibility and new opportunities. While the M-24-10 Memorandum is explicitly related to government-developed AI, private sector companies face the same issues called out in section 5 B and C \cite{whitehouse2410} around testing real-world scenarios. Without established certification criteria, businesses are forced to operate in a vacuum needing to prove their services work, and accepting any liability from unstable regulation. 

For certifying authorities, it offers a path to consumer trust in AI, and a process to stay up-to-date with risks in each relevant use case. The need for established guidelines around AI is directly referenced in Section 4.1 of the EO 14110 \cite{exec14110}, and the NIST 600-1 Generative Profile \cite{nistai} references the need to develop certification programs relevant to specific industry and context.

\subsection{Dataset Curation}

Curating datasets designed to evaluate and certify domain-specific consumer facing experiences is the starting point for a robust LLM-based experience certification process. As model performance continues to improve, high quality data has become the scarcest available resource \cite{shumailov2024collapse}.

These datasets should contain sample user input relevant to specific use cases, with example ground truth responses accompanying each input. This will require subject matter expertise, as well as constant dataset management. Depending on the use case, this might also include multi-modal data as it becomes more commonplace in LLM-based solutions. These datasets should be dynamic, responding to changes in the marketplace, as well as consumer and business partner feedback.

This paper does not aim to trivialize this effort. Creating and maintaining these datasets will be labor intensive, and means to make it more efficient like pursuing synthetic data are advisable \cite{raju2024judge, guo2024synthetic}. 

\subsection{Certification Focus Areas}

Given the generalized nature of LLMs, it is not feasible to provide a certification process for every possible use case, so use cases will need to be prioritized based on perceived societal benefit. The EU AI Act \cite{euact} outlines categories of risk. In the United States, Appendix 1 Sections 1 and 2 from the Whitehouse M-24-10 Memorandum \cite{whitehouse2410} highlight safety and rights impacting use cases, such as providing medical diagnoses and detecting student plagiarism. The purpose of the memorandum is to manage risk for usage of AI by the Federal government, but it is a logical extension to use this list as a starting point for a potential certification process for the private sector. While evaluations should be domain specific, a single LLM-based experience could obtain one or many certifications across domains.

\subsection{Example Data Curation and Certification Workflow}

The example below aims to make a potential version of the certification process described in the paper more tangible.

\begin{enumerate}
    \item Work with internal and industry experts to create relevant prompts for a specific use case.
    \item Experts evaluate prompts to produce a ground truth dataset of how to respond to each prompt. This becomes the prompt-response dataset used in evaluation. LLMs-as-judge \cite{raju2024judge, zheng2024judge, murugadoss2024Evaluating} could be used to extend the evaluation process for efficiency purposes if deemed acceptable. In conjunction with this work, a rubric should be developed to assist expert evaluation. The means of evaluation could be binary, on a Likert scoring system, or some other equivalent.
    \item A passing criteria for certification submissions should be established leveraging the rubric developed and chosen scoring system. Passing could ultimately be represented in tiers.
    \item Share a training set with the public to give scope of expected evaluation. Keep a test set hidden to perform the certification evaluation. This test set should be continually curated as a means to prevent the evaluation from directly leaking into model training.
   \item A phased approach for auditing and evaluation.
    \begin{enumerate}
        \item Phase 1 - have experts manually review output to determine submission outcome.
        \item Phase 2 - leverage the ground truth dataset to convert evaluation into a model-based, automated process. There should always be a manual review for auditing purposes, but automated evaluation would allow for significantly more scalability of the process.
    \end{enumerate}
\end{enumerate}

\section{Discussion and Recommendations}

Pivoting to dataset-driven LLM regulation requires a policy shift to step away from treating all LLM-powered AI systems as one uniform entity, to targeting specific use cases that can drive business growth and significant consumer value, without regulatory overreach that stifles innovation.

As the AI community continues to make advances, it will become harder to regulate at the compute-level. New techniques may make current compute thresholds obsolete. LLMs can also be distributed in model weight form for downstream consumption and alteration, making it unrealistic to comprehensively control these models in a preventative fashion at the model training level. 

Data itself should be the strategic center of any regulatory strategy moving forward. High quality labeled data has become a significant bottleneck in the LLM training and evaluation processes, leading to an emphasis on synthetic data generation \cite{raju2024judge, liu2024synthetic}. Generating and maintaining high quality datasets positions regulatory bodies well for a landscape where algorithms and knowledge are largely commoditized, and a means to evaluate AI systems is the most critical form of consumer protection.

We hope this paper can stimulate conversation around using data and evaluation as the primary means of LLM and AI regulation. We believe this will drive regulation to a more productive place that benefits consumers, businesses and government entities alike.

\bibliographystyle{IEEEtran}
\bibliography{references.bib}

\section{Appendix A: Proposed Certification Logistics}

When a company successfully completes an LLM certification process, they will receive official documentation they can display in their marketing materials, documentation, and communications with users. These companies will need to make clear to users that their experience is powered by an LLM to ensure no confusion.

While each certification will be targeted to a specific use case, a single LLM-based experience can apply to multiple different certification processes. Given the generalized nature of the technology, it's feasible that a single company or even single experience could qualify for multiple of identified use cases, such as being a mental health coach and a dietary coach.

The timing associated with re-certification, and system update triggers that might necessitate re-certification should be explicitly defined as part of the certification process.

\end{document}